\documentstyle[aps,prl,floats,epsf]{revtex}

\def\Ce8{Ce$_{0.8}$La$_{0.2}$Al$_3$}
\def\Jpar{$J_\parallel$}
\def\Jperp{$J_\perp$}
\def\Cpar{$\chi_\parallel$}
\def\Cperp{$\chi_\perp$}

\begin{document}
\draft
\wideabs{

\title{Evidence for Anisotropic Kondo Behavior in Ce$_{0.8}$La$_{0.2}$Al$_3$}

\author{
         E. A. Goremychkin,$^1$
         R. Osborn,$^1$
         B. D. Rainford,$^2$
         and A. P. Murani$^3$
        }

\address{ $^1$ Argonne National Laboratory, 9700 South Cass Avenue,
          Argonne, IL 60439 }

\address{ $^2$ Department of Physics, University of Southampton,
          Southampton SO17 1BJ, United Kingdom }

\address{ $^3$ Institut Laue-Langevin, BP 156, F-38042 Grenoble
           C\'{e}dex 9, France }

\date{\today}

\maketitle

\begin{abstract}
We have performed an inelastic neutron scattering study of the low energy
spin dynamics of the heavy fermion compound Ce$_{0.8}$La$_{0.2}$Al$_3$ as
a function of temperature and external pressure up to 5 kbar.  At
temperatures below 3 K, the magnetic response transforms from a quasi-elastic
form, common to many heavy fermion systems, to a single well-defined
inelastic peak, which is extremely sensitive to external pressure.
The scaling of the spin dynamics and the thermodynamic properties are
in agreement with the predictions of the anisotropic Kondo model.
\end{abstract}

} % end \wideabs

CeAl$_3$ was the first material to be classified as a heavy fermion system
over twenty years ago \cite{Andres75}.  At first, it was thought to have
a non-magnetic ground state, and anomalous features in the thermodynamic and
transport properties at about $T^*$ $\sim$ 0.5 K were interpreted as the
signature of a transition from a single-ion Kondo regime to a coherent Kondo
lattice regime \cite{Bredl84}.  However, Barth {\it et al.} \cite{Barth89}
presented evidence from muon spin relaxation of quasi-static internal
magnetic fields at temperatures below 0.7 K, which suggested the existence
of frustrated short-range magnetic order.  Since then,
there has been conflicting evidence concerning the existence of magnetic
order in CeAl$_3$ \cite{Wong92}. Recently, Andraka {\it et al.} \cite{Andraka95}
reported that the temperature of the anomalies in the specific heat and
magnetic susceptibility increases gradually with lanthanum doping,
from $T^*$ $\sim$ 0.5 K in pure CeAl$_3$ to $T^*$ $\sim$ 2.2 K in
\Ce8, and interpreted this as evidence of a stabilization of the\
antiferromagnetic ground state due to the changing balance of Kondo
and RKKY interactions.

The aim of our investigation was to clarify the microscopic origin of the
low temperature anomalies in Ce$_{1-x}$La$_x$Al$_3$ through measurements
of the dynamic magnetic correlations, using neutron scattering and muon spin
relaxation ($\mu$SR).  We find a dramatic evolution of the dynamic magnetic
susceptibility, from a quasi-elastic response at high temperature to an
inelastic response below $T^*$. In \Ce8, the inelastic peak is at
approximately 0.5 meV at 1.7 K. It is extremely sensitive to applied pressure
and the dynamics become purely relaxational above only 2 kbars.  Although we
observe a sharp increase in the $\mu$SR relaxation rate at $T^*$, suggesting
the development of static magnetic correlations, we find no evidence of
long-range magnetic ordering from high-intensity neutron diffraction.  We
argue that the anomalous response of \Ce8\ is consistent with the predictions
of the anisotropic Kondo model \cite{Chakravarty95,Costi96,Costi98}, which has
been attracting considerable theoretical attention recently because of its
relation to statistical models of two-level systems with Ohmic
dissipation \cite{Leggett87}.  In this interpretation, the transition at $T^*$
is driven by the onset of weakly-dissipative single-ion dynamics, rather than
cooperative magnetic ordering.

The sample of \Ce8\ was prepared by arc melting stoichiometric quantities of
the constituent elements, followed by annealing at 900$^\circ$ C for about
four weeks.  Both neutron and x-ray diffraction confirmed that the sample was
single phase.  The neutron scattering experiments were performed at the
Institut Laue-Langevin on the time-of-flight spectrometer IN6, using an
incident energy of 3.1 meV, the high flux powder-diffractometer D20,
using a wavelength of 2.41 \AA, and the polarized diffractometer D7,
using a wavelength of 4.8 \AA.  A continuously loaded helium high-pressure
cell, operating up to a maximum pressure of 5 kbar, was inserted into a
standard helium cryostat for the IN6 experiments.  The zero-field $\mu$SR
measurements were performed at the ISIS pulsed muon facility, using the
MUSR spectrometer.  

\begin{figure}
 \centering
 \epsfxsize=8.4cm
 \epsfbox{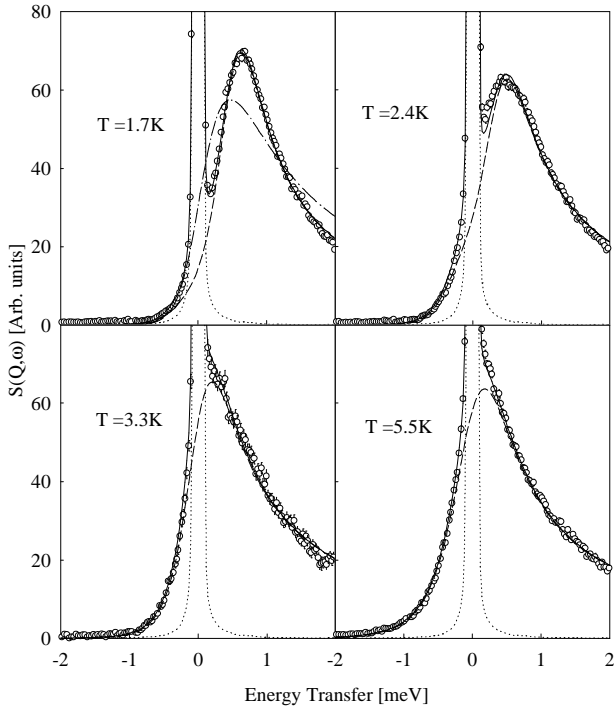}
 \vspace{0.3cm}
 \caption{ Inelastic neutron scattering data (open circles) from \Ce8\ 
 measured on the IN6 spectrometer with an incident energy of 3.15meV.
 The spectra were summed over momentum transfers from 0.3 to 1.1
 \AA$^{-1}$. The dotted line is the elastic nuclear scattering,
 the dashed line is the profile fit of the
 Eqn.\ (\protect\ref{SQw}), and the solid line is the sum of both
 contributions.  The dash-dotted line is the best fit at T = 1.7K to a
 quasielastic lineshape.
 \label{Fig1} }
\end{figure}

An example of low temperature IN6 data, in the form of the scattering law
$S(Q,\epsilon)$ is shown in Fig.\ \ref{Fig1}. We have modeled
$S(Q,\epsilon)$ using a standard Lorentzian lineshape
\begin{eqnarray}
 \label{SQw}
  S(Q,\epsilon) \propto
  && F^2(Q) \frac{\epsilon \chi_0}
      {1 - \exp(-\epsilon/kT)} \nonumber \\
  && \times \frac{1}{2\pi} \left(
   \frac{\Gamma}{\left[ (\epsilon - \Delta)^2 + \Gamma^2 \right]} +
   \frac{\Gamma}{\left[ (\epsilon + \Delta)^2 + \Gamma^2 \right]}
  \right)
\end{eqnarray}
where $F(Q)$ is the Ce$^{3+}$ magnetic form factor, $\chi_0$ is the static
susceptibility, and $\Gamma$ the half width at half-maximum of the
Lorentzians centered at energy transfers $\pm\Delta$. The usual paramagnetic
response of heavy fermion compounds, including CeAl$_3$ \cite{Murani80},
is purely relaxational (i.e.\ $\Delta$ = 0), typically with a square
root temperature dependence of the linewidth:
$\Gamma(T) = \Gamma_0 + \beta \sqrt{T}$ \cite{Horn81}.
Our analysis of the IN6 data from \Ce8\ shows that the low energy spin
dynamics follow this behavior from 3.3 K up to 100 K with $\beta$ =
0.18(1) meVK$^{-1/2}$ and $\Gamma_0$ = 0.34(3) meV.  However, there is a
radical change in the shape of response function below 3.3 K, from a
quasi-elastic to an inelastic form (i.e.\ $\Delta$ $\neq$ 0).
Figure \ref{Fig1} shows that the fit to a quasi-elastic response gives a
poor description of the data at 1.7 K.  The peak energy, which is at
0.474(7) meV at 1.7 K, is weakly temperature dependent; we estimate that
it increases to 0.54(1) meV at zero temperature.
The linewidth is 0.42(1) meV.

We note that there is no possibility that this inelastic peak results from
a magnon-like excitation within the ground-state doublet.  The crystal field
(CF) potential in \Ce8\ is a polynomial in $(J_z)^2$, producing a ground
state doublet $\Gamma_9$ $(|\pm 3/2\rangle)$ and two excited doublets
$\Gamma_7$ $(|\pm 1/2\rangle)$ and $\Gamma_8$ $(|\pm5/2\rangle)$ at an energy
of 7.4 meV \cite{Goremychkin99}.  There is no dipole matrix element coupling
the $|\pm 3/2\rangle$ states, so conventional magnons would not be measurable
with neutrons \cite{Rainford96}. Moreover, the excitation energies and
linewidths are only weakly $Q$-dependent, indicating that interionic exchange
interactions are small.  Nevertheless, the development of this inelastic peak
approximately coincides in temperature with peaks in both the specific heat
and bulk susceptibility, so we have performed both $\mu$SR and neutron
diffraction measurements to look for evidence of magnetic ordering below 2 K.

\begin{figure}
 \centering
 \epsfxsize=8.4cm
 \epsfbox{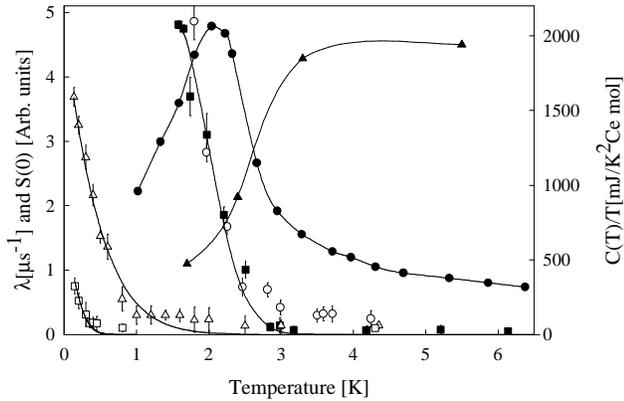}
 \vspace{0.3cm}
 \caption{ Temperature dependence of (1) the $\mu$SR relaxation rate
 $\lambda$ for Ce$_x$La$_{1-x}$Al$_3$:
 $x$ = 0.8 (solid squares; absolute values divided by 3);
 $x$ = 0.5 (open circles); $x$ = 0.3 (open triangles),
 and $x$ = 0.1 (open squares),
 (2) the neutron scattering function at zero energy transfer,
 $S(\epsilon\rightarrow 0)$ (solid triangles),
 and (3) the specific heat $C(T)/T$ (solid circles; from
 Ref.\ \protect\cite{Andraka95}).  The lines are guides to the eye.
 \label{Fig2} }
\end{figure}

Zero-field $\mu$SR spectra were measured on samples of Ce$_x$La$_{1-x}$Al$_3$
with $x$ = 0.8, 0.5, 0.3 and 0.1. All spectra could be fitted to the function
$G(t)$, comprising the sum of a  Lorentzian and a Kubo-Toyabe (KT)
depolarization function, $G(t) = A_1 \exp(-\lambda t) + A_2 G_{\text{KT}}(t)$,
where $G_{\text{KT}}$ accounts for muon precession due to the dipole fields
of the $^{27}$Al nuclear magnetic moments \cite{Barth89,Amato97}.
Lorentzian damping usually arises from dynamical processes, but here
(as in CeAl$_3$ \cite{Amato97}), it is likely to arise from
inhomogeneous broadening due to a static, or quasi-static, distribution of
fields arising from magnetic ordering of the 4f electrons.  There was no
evidence, within our limited time resolution, for a second Kubo-Toyabe
function or an oscillatory component, as seen in the $\mu$SR data for
CeAl$_3$ \cite{Barth89,Amato97}.  At high temperatures, the spectra are
mainly determined by the relaxation of the $^{27}$Al nuclear moments.
Below a characteristic temperature, $T^*$ ($\sim$ 3K for
$x$ = 0.8 and 0.5, $\sim$ 1.5 K for $x$ = 0.3, and $\sim$ 0.5K for $x$ = 0.1),
the Lorentzian damping starts to contribute to $G(t)$ (see Fig.\ \ref{Fig2}),
with a sharp increase in the damping rate as the temperature is lowered
below $T^*$.  The temperature at which $\lambda$ diverges
corresponds to the maximum in the specific heat for $x$ = 0.8
(see Fig.\ \ref{Fig2}).

In support of our attribution of the Lorentzian damping in the $\mu$SR data
to inhomogeneous broadening arising from quasi-static fields, we note that
the increase in the $\mu$SR relaxation rate at $T^*$ occurs when the
response becomes purely inelastic.   If the damping rate $\lambda$ were
attributed to dynamical processes, we would expect
$\lambda$ $\approx$ $\lim_{\epsilon\rightarrow 0} T\chi^{\prime\prime}
(\epsilon)/\epsilon \approx S(\epsilon\rightarrow 0)$.
It follows that we would expect $\lambda$ to {\it decrease}  below $T^*$,
since $S(0)$, determined from the analysis of the 
inelastic lineshape,
falls dramatically at $T^*$ (see Fig.\ \ref{Fig2}).  On the other hand
an increase in $\lambda$ should give rise to an increase in low-frequency
magnetic response measured by neutron scattering, for which there is no
evidence in the IN6 data. There is no increase in the elastic peak
intensity and therefore no evidence of a transfer of spectral weight to
an unidentified low-frequency component.

The most direct method of determining the presence of long-range magnetic
ordering is neutron diffraction.  We have performed a series of experiments
on $x$ = 0.8 and 0.5 samples on the high-intensity powder diffractometer D20.
The difference in the diffraction patterns measured below and above $T^*$
shows no evidence of any magnetic Bragg peaks in both compounds.
We estimate the Ce magnetic moment in any magnetically ordered phase of
\Ce8\ to be less than 0.05 $\mu_B$. Furthermore, measurements 
on D7 with full polarization analysis gave no sign of either
magnetic Bragg peaks or significant short-range magnetic order at 1.5 K.
It should be noted that the estimated value of the Ce magnetic moment
from the heat capacity anomaly of \Ce8\ is 0.34 $\mu_B$ \cite{Andraka95},
which would easily be seen by neutron diffraction.  However we argue
below that most of this heat capacity anomaly arises from the change in
form of the single ion dynamics at $T^*$, so that this estimate of the
ordered moment may be discounted.
The evidence from the neutron diffraction experiments suggests
that if magnetic ordering occurs below $T^*$, as suggested by the $\mu$SR
data, it must be associated with an extremely small amplitude of the
magnetic moment.  We can then infer that the exchange field will be far
too small to drive the change in the dynamical response below $T^*$.
In order to find an explanation that can encompass the entire composition
range from $x$ = 0.0 to 0.9, we have explored an alternative account of
the dynamical behavior.

In recent years, there has been substantial interest in the anisotropic
Kondo model (AKM) \cite{Chakravarty95,Costi96,Costi98} as a theoretical
realization of a dissipative two-state system (DTSS) \cite{Leggett87}.
In the AKM, the interaction between a localized spin and the conduction
electrons is anisotropic and there are two parameters representing
the strength of this interaction, \Jpar\ and \Jperp, with \Jpar\ $\neq$ \Jperp.
CeAl$_3$ is a natural candidate for such a model, because the anisotropy of
the low-temperature magnetic susceptibility is very strong;
the $|\pm3/2\rangle$ crystal field ground state is Ising-like
($g_\perp$ = 0 and $g_\parallel$ = 18/7) and the bulk susceptibility is
dominated by a divergent Curie term \Cpar(T) below 40 K \cite{Goremychkin99}.
The Van Vleck contribution from the excited crystal field states, \Cperp(T),
is much weaker.

The dynamics of the DTSS model are governed by the bare splitting of the two 
states at an energy $\Delta_0$, and the strength of the dissipation produced
by the bosonic continuum, which is characterized by the dimensionless
parameter $\alpha$.  Below a critical value of $\alpha$, estimated to be
roughly 1/3 by Costi and Kieffer \cite{Costi96}, the dynamic response is
predicted to be inelastic, but when $\alpha$ $>$ 1/3, the response is
overdamped.
The DTSS model may be mapped onto the AKM, with \Jperp\ proportional to
$\Delta$ and \Jpar\ associated with $\alpha$.  In the weakly-dissipative
regime of interest to us, i.e., when $\alpha$ $<$ 1/3,  \Jpar\ $\gg$
\Jperp.

It is possible to obtain independent estimates of the value of $\alpha$
from the specific heat and magnetic susceptibility results, using scaling
relations predicted by the numerical calculations.  Firstly, the theory
predicts that there is a peak in $C(T)/T$, at a temperature $T^*$ =
$\alpha/\gamma$, where $C(T)$ is the specific heat and $\gamma$ is the
value of $C(T)/T$ for $T$ $\ll$ $T^*$ \cite{Costi98}.  The specific
heat has only been measured down to 1 K in \Ce8, but we estimate that
$\gamma$ should be in the range 0.4-0.5 Jmol$^{-1}$K$^{-2}$ from the
extrapolated values measured for 0.9 $<$ $x$ $<$ 1.0 \cite{Andraka95}.
Since $T^*$ $\approx$ 2K \cite{Andraka95}, we obtain $\alpha$ = 0.10
$\pm$ 0.02.  As a check on this result, we note that $\alpha$ is also
given by the inverse Wilson ratio $\gamma/\chi$.  From $\chi(T=1.8K)$
= 0.03 emu/mol \cite{Andraka95}, we obtain the identical value
$\alpha$ $\approx$ 0.10.  This represents the upper limit of $\alpha$,
because we are likely to have overestimated $\gamma$.

This value of $\alpha$ falls in the regime where the AKM predicts
an inelastic response.  There have been several numerical calculations of
$S(\epsilon)$, which show that it peaks at a renormalized energy $\Delta$,
which scales as $\Delta_0(\Delta_0/\omega_c)^{\alpha/(\alpha-1)}$ where
$\omega_c$ is the conduction electron bandwidth.  Combining the AKM prediction for
the bulk susceptibility, $\chi(T=0)$ = $\mu_B^2N_A/2\Delta$ with the
measured value of 0.03 emu/mol, gives $\Delta$ $\sim$ 0.54 meV.
Furthermore, the AKM predicts that $\gamma/\alpha$ = $\pi^2k_B^2/3\Delta$
\cite{Costi98}, from which we estimate that $\Delta$ is between 0.47
and 0.59 meV.  The predicted values for $\Delta$ are consistent
with the energy of the inelastic peak measured by neutron scattering.

\begin{figure}
 \centering
 \epsfxsize=8.4cm
 \epsfbox{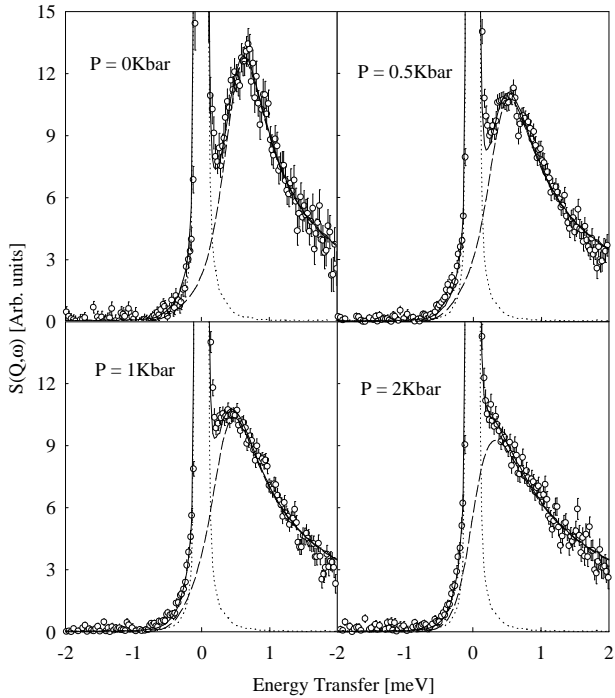}
 \vspace{0.3cm}
 \caption{Inelastic neutron scattering data from \Ce8
 vs.\ pressure at $T$ = 1.7K.  The symbols and lines are the same as in
 Fig.\ \protect\ref{Fig1}.
 \label{Fig3} }
\end{figure}

If we apply the same arguments to pure CeAl$_3$, we find that
$\alpha$ $\approx$ 0.31 (using $\gamma$ = $C(T=50 \text{mK})/T$ = 1.35
Jmol$^{-1}$K$^{-2}$ \cite{Brodale86} and $\chi(T=40 \text{mK})$ =
0.0295 emu/mol \cite{Avenel92}), which is very close to the critical value
of $\alpha$ = 1/3 where the response function $S(\epsilon$) become
quasi-elastic.  Inelastic neutron studies of CeAl$_3$ have shown
that the magnetic dynamics remain quasi-elastic down to 60 mK \cite{Murani80}.
According to the AKM, the decrease in $\alpha$ with increasing $x$ would be
due to the ``negative" chemical pressure produced by lanthanum dilution.
To test this, we have measured the effect of ``positive" external hydrostatic
pressure on the magnetic response of \Ce8\ to see if it drives the system
closer to pure CeAl$_3$.  Figure \ref{Fig3} shows that the effect of external
pressure is remarkably strong, with a reduction in the peak energy evident
at only $P$ = 0.5 kbar.  At 2 kbar, the magnetic response is once again
quasi-elastic. $\Delta$ has an almost linear dependence on pressure with
$d\Delta/dP$ = -0.24 meV/kbar at 1.7 K. We only observe such a strong
pressure dependence in the vicinity of $T^*$;
at 5 K, the magnetic response is practically pressure-independent.
The effects of pressure indicate that the differences in the dynamical
behavior of CeAl$_3$ and \Ce8\ are not the result of chemical disorder.
In the framework of the AKM, it means that the dissipation strength 
$\alpha$ is extremely pressure-dependent.  The observation that the
specific heat of pure CeAl$_3$ is only pressure-dependent close to
the maximum in $C(T)/T$ \cite{Brodale86} is consistent with this explanation.

We have argued that, as in CeAl$_3$, the $\mu$SR results are evidence of
magnetic ordering, of either short or long range, with moments less than 
$\sim$ 0.05 $\mu_B$, similar to what has been observed in several uranium
heavy fermion compounds \cite{Amato97}.  The spectral weight associated
with such weak magnetic correlations would not be measurable by neutron
scattering from polycrystalline samples.  Costi and Kieffer \cite{Costi96}
showed that the inelastic peak in $S(\epsilon)$ persists in the presence
of a bias field, as long as this is small compared to $\Delta$.
Equating the bias field with the internal exchange field, we infer that
antiferromagnetism or spin glass order with weak moments would be unlikely
to change the form of the single-ion dynamics.  Such ordering would be
insufficient to drive the thermodynamics and so is more likely to be a
by-product of the reduced dissipation of the ground state.  In this scenario,
the transition at $T^*$ is produced by the single-ion AKM, but allows the
development of a more coherent $f$-electron band at low temperature. 
The small-moment magnetism would be a manifestation of the itinerant nature
of the $f$-electrons in this regime.

Other non-cubic heavy fermion systems, such as CeRu$_2$Si$_2$, have highly
anisotropic susceptibilities, so they should be considered
in the context of the anisotropic Kondo effect; indeed, the heuristic model
used to describe the spin dynamics in CeRu$_2$Si$_{2-x}$Ge$_x$alloys
\cite{Rainford96} had many of the features of the dynamics of the AKM
\cite{Costi96}.  The unusual properties of URu$_2$Si$_2$, which is also 
highly anisotropic, might be associated with the AKM.  Most of
the spectral weight is in the dynamical response, which is also characterized
by longitudinal fluctuations.  The ordered moment is very small
(0.04 $\mu_B$), yet there is a large heat capacity anomaly at
$T_N$ \cite{Broholm91,Palstra85}.  An extension
of the AKM to account for intersite interactions would be necessary for a
description of the strong dispersion of the magnetic excitations.

In conclusion, we have shown that anomalies in the specific heat and magnetic
susceptibility of \Ce8\ are not driven by the development of static magnetic
correlations, but are associated with the development of the single-ion
inelastic response of the cerium 4$f$-electrons, arising from their coupling
to the conduction electrons. The scaling of the bulk and dynamic properties
is consistent with the predictions of the anisotropic Kondo model, from which
we conclude that the anisotropy of coupling in CeAl$_3$ is close to the
critical value at which a weakly-dissipative response is observable.
These results provide a new insight into the mechanisms by which 
low-temperature coherence is established in heavy fermions.

%\acknowledgments
We thank I.\ Sashin, T.\ Hansen, C.\ A.\ Scott, and K.\ H.\ Andersen for
their help with the experiments, and D.\ Grempel for useful discussions.  EAG 
thanks the Rutherford Appleton Laboratory for hospitality and financial 
assistance.  This work was supported by the 
US Department of Energy Contract No.\ W-31-109-ENG-38.

\end{document}